\documentclass[aps,twocolumn,tightenlines,showpacs]{revtex4}
\usepackage[dvipdf]{graphicx}
\usepackage{color}

\begin{document}
\title{He-McKellar-Wilkens topological phase in atom interferometry}

\author{S.~Lepoutre, A.~Gauguet, G.~Tr\'enec, M.~B\"uchner, and J.~Vigu\'e}
\address{ Laboratoire Collisions Agr\'egats R\'eactivit\'e-IRSAMC
\\Universit\'e de Toulouse-UPS and CNRS UMR 5589, Toulouse, France
\\ e-mail:~{\tt jacques.vigue@irsamc.ups-tlse.fr}}

\date{\today}

\begin{abstract}

We report the first experimental test of the topological phase predicted by He and McKellar and by Wilkens in 1993: this phase, which appears when an electric dipole propagates in a magnetic field, is connected to the Aharonov-Casher effect by electric-magnetic duality. The He-McKellar-Wilkens phase is quite small, at most $27$ mrad in our experiment, and this experiment requires the high phase sensitivity of our atom interferometer with spatially separated arms as well as symmetry reversals such as the direction of the electric and magnetic fields. The measured value of the He-McKellar-Wilkens phase differs by $31$\% from its theoretical value, a difference possibly due to some as yet uncontrolled systematic errors.
\end{abstract}

\maketitle
\bigskip

In quantum mechanics, propagation can be modified without any force,
the first example being the Aharonov-Bohm effect \cite{aharonov59}
discovered in 1959: a magnetic field shifts the fringes of an
electron interferometer, even if the field vanishes on the electron
paths. This was the discovery of topological phases \cite{shapere89},
which differ considerably from ordinary dynamic phases because they
are independent of the particle velocity and non-reciprocal, i.e.
they change sign when propagation is reversed. In 1984, Aharonov and
Casher \cite{aharonov84} discovered another topological phase, which
appears in a matter wave interferometer operated with a particle
carrying a magnetic dipole, the interferometer arms encircling a line
of electric charges. In 1993, He and McKellar \cite{he93} applied
electric-magnetic duality \cite{lepoutreSM} to the Aharonov-Casher
phase, thus exhibiting a topological phase when the particle carries
an electric dipole and the interferometer arms encircle a line of
magnetic monopoles: this phase appeared as speculative but a possible
experiment was rapidly proposed by Wilkens \cite{wilkens94}. Whereas
the Aharonov-Bohm and Aharonov-Casher (AC) effects were rapidly
tested by experiments
\cite{chambers60,tonomura86,cimmino89,sangster93,gorlitz95,zeiske95},
no experimental test of the He-McKellar-Wilkens (HMW) phase has been
available so far. Here, we report the first experimental attempt to detect 
the HMW phase, with results in reasonable agreement with theory.
The HMW phase is the last member of the family of topological phases
when free particles propagate in electromagnetic fields
\cite{muller95,dowling99}: one might expect similar phases for higher
order electromagnetic multipoles  but the calculated values for
quadrupoles \cite{chen95} are so small that their detection is
presently out of reach.

Although Wilkens experiment proposal \cite{wilkens94} is 18 years
old, no experimental proof of the HMW phase has been reported. A test
was discussed in 1996 by Wei {\it et al.} \cite{wei95} and by
Schmiedmayer {\it et al.} \cite{schmiedmayer97} and an experiment
with a superfluid helium interferometer was proposed in 2009 by Sato
and Packard \cite{sato09}, without any published result yet. Let us
compare the detection of the AC and HMW phases, in order to
understand why this test is difficult. All accurate tests of the AC
phase \cite{sangster93,gorlitz95,zeiske95} have used a Ramsey
interferometer \cite{ramsey50}, in which the particle propagates in a
superposition of two spin states: this type of interferometer is
ideal for the detection of a spin-dependent phase and it provides an
excellent cancelation of systematic errors. The use of a Ramsey
interferometer for the HMW phase would require the production of a
quantum superposition of states with opposite electric dipole
moments, which is feasible if states of opposite parity are
quasi-degenerate \cite{dowling99}, a situation which does not exist
with ground state atoms. Consequently, the HMW phase must be measured
by alternating field configurations and by studying differences of
measured phases: this procedure makes an experiment more subject to
systematic errors than Ramsey interferometry. In addition, in order
to create a non-zero HMW phase shift, the two interferometer arms
must propagate in different electric or magnetic fields, which is
possible only if the two arms are spatially separated. Following the
pioneering work of Pritchard and co-workers
\cite{keith91,schmiedmayer97}, very few separated-arm atom
interferometers have been built.

Although not associated with a classical force, the AC and the HMW
phases can be explained by the interaction of a dipole with a
motional field \cite{dowling99}. The HMW phase is due to the
interaction of an electric dipole $\mathbf{d}$ with the motional
electric field $\mathbf{E}_{mot}= \gamma \mathbf{v} \times
\mathbf{B}$ where $\mathbf{v}$ is the atom velocity, $\mathbf{B}$ is
the magnetic field and $\gamma= 1/\sqrt{1-v^2/c^2} \simeq 1$ is the
relativistic factor. $U=-\mathbf{d}\cdot \mathbf{E}_{mot}$ is the
interaction energy, which induces the HMW phase shift:

\begin{equation} \label{e1}
\varphi_{HMW}= \oint U dt/\hbar
\end{equation}

\noindent where the closed loop follows the interferometer paths.
Because of parity, atoms or molecules do not have a permanent dipole
and an electric field $\mathbf{E}$ is needed to induce a dipole
$\mathbf{d}= 4\pi\varepsilon_0 \alpha \mathbf{E}$, where $\alpha$ is
the electric polarizability. Finally, the HMW phase shift is given
by:

\begin{equation}\label{e2}
\varphi_{HMW}=  \oint \left(\mathbf{d} \times \mathbf{B}\right) \cdot \mathbf{v} dt/\hbar
\end{equation}

\noindent Because $\mathbf{v} dt$  is the infinitesimal path length,
$\varphi_{HMW}$  is independent of the modulus $v$ of the atom
velocity but it changes sign with the direction of propagation, and
it is maximum when $\mathbf{E}$, $\mathbf{B}$ and $\mathbf{v}$ are
orthogonal.

\begin{figure}[h]
\begin{center}
\includegraphics[height= 7 cm]{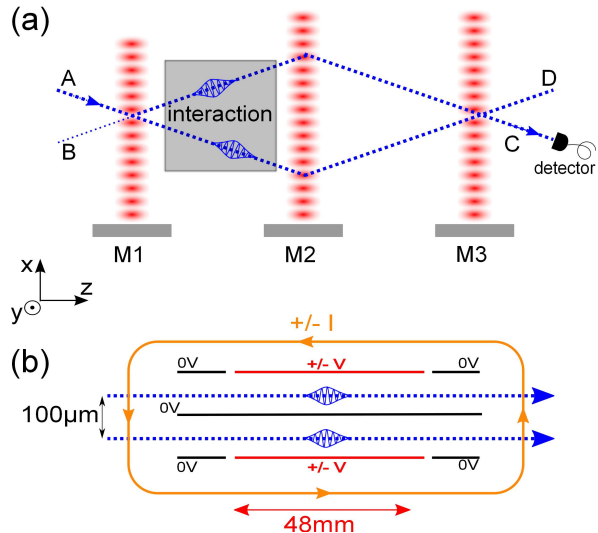}
\caption{ (Color online) Schematic top views of the  setup. Panel a: our atom interferometer, with two entrances A and B  and two exits C and D (C is detected). An atomic beam (dotted blue lines) entering by A is diffracted by
three quasi-resonant laser standing waves produced by the mirrors M$_i$. The interaction region is placed  where the distance between interferometer arms is largest, close to $100$ $\mu$m. Panel b: the interaction region producing the electric and magnetic fields (not to scale). The interferometer arms (dotted blue lines) are separated by a septum, which is the common electrode of two plane capacitors producing opposite electric fields (high voltage electrodes in red; grounded electrodes in black). Two
rectangular coils (brown rectangle) produce the magnetic field. \label{fig1}}
\end{center}
\end{figure}

To detect the HMW phase, we use a Mach-Zehnder atom interferometer
\cite{miffre05b} shown in fig. \ref{fig1}a. A supersonic
lithium beam (mean velocity near 1000 m/s with a distribution half-width close to $100$ m/s) is strongly collimated and then
diffracted by three laser standing waves in the Bragg regime: the
laser frequency is chosen such that only the dominant lithium isotope
$^7$Li contributes to the signal \cite{miffre05b,jacquey07}. The
diffraction events play the roles of beam-splitters and mirrors for
the atomic wave. This interferometer produces two output beams,
labeled C and D on fig. \ref{fig1}a, with complementary fringe
signals. Beam C is selected by a slit and the atoms
of this beam are ionized by a Langmuir-Taylor "hot-wire" detector  \cite{delhuille02b}.
The resulting ions are counted, thus providing the interferometer
signal $I$ given by:

\begin{equation} \label{e3}
I=I_0[1+ \mathcal{V} \cos(\varphi_p + \varphi_d)]
\end{equation}

\begin{figure}[h]
\begin{center}
\includegraphics[height= 6.75 cm]{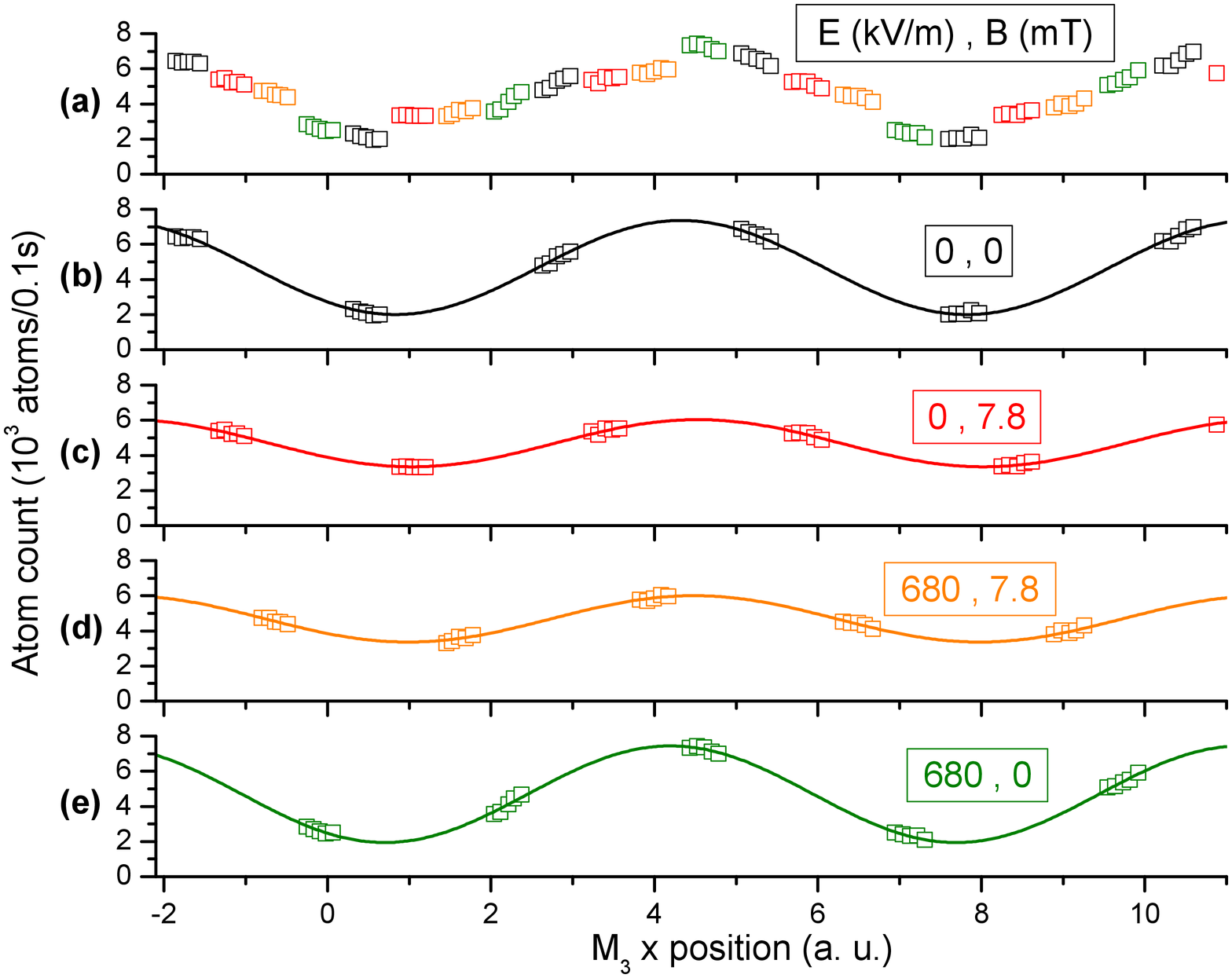}
\caption{(Color online) Panel a: the atom interferometer signal is
plotted as a function of M$_3$ $x$-position, using the 4-field
configuration procedure with $E = 680$ kV/m and $B = 7.8$ mT. In
panels b-e, the signals corresponding to each field configuration,
are plotted with their best fits. The mean detected flux
$I_0\approx47 \times 10^3$ atoms/s is constant. We use the no-field
case as reference for visibility ($\mathcal{V}_0 = \left( 57.2 \pm
0.6 \right)$ \%) and phase. In panel c, $\mathcal{V} \left(0, I
\right)/\mathcal{V}_0 = \left(50.0\pm 1.2\right)$\% and $\varphi_B
\left(I \right) = - 183 \pm 25 $ mrad. In panel d, $\mathcal{V}
\left(V, I \right)/\mathcal{V}_0 = \left(48.7 \pm 0.2\right)$\% and
$\varphi_{E+B}\left(V, I \right) = -139 \pm 24 $ mrad.  In panel e,
$\mathcal{V} \left(V, 0 \right)/\mathcal{V}_0 = \left(102.4 \pm
1.6\right)$\% and $\varphi_E \left(V \right) = 118 \pm 18 $ mrad.
>From these measurements, we get $\varphi_{EB} \left(V, I \right)=
\left(-74 \pm 35\right)$ mrad. The average over about 100 similar
scans produce phase measurements with a few mrad error bar.
\label{fig2}}
\end{center}
\end{figure}

\noindent  $I_0$ is the mean intensity and $\mathcal{V}$ is the
fringe visibility. The phase $\varphi_p$ is due to perturbations. The
phase $\varphi_d$, due to atom diffraction, depends on the positions
of the laser standing wave mirrors M$_i$. To record interference
fringes (see fig. \ref{fig2}), we move mirror M$_3$ with a
piezoelectric device and its displacement, measured by a Michelson
interferometer, gives very accurately the variations of  $\varphi_d$.
Typical values of the mean intensity and of the fringe visibility are
$I_0 \approx 50000$ atoms/s and $\mathcal{V}\approx 70$\%, leading to
a phase sensitivity near $20-30$ mrad/$\sqrt{\mbox{Hz}}$.

The interaction region is schematically represented in fig.
\ref{fig1}b (more details in \cite{lepoutreSM}): homogeneous electric
fields are produced by two plane capacitors sharing a thin "septum"
electrode \cite{ekstrom95}, which is inserted between the two
interferometer arms without modifying their propagation. With a
$1.10$ mm electrode spacing, the electric field is $E$(kV/m)$\approx
0.9 V$ where the applied voltage $V$ (Volt) can be positive or
negative, in order to test field reversal. The double capacitor is
placed inside a pair of coils producing a fairly homogeneous magnetic
field: the use of coils rather than permanent magnets limits the
field value near $14$ mT but gives rise to a much better control and
permits field reversal. For a current $I$ circulating in the coils,
the field $B$ at the center is given by $B/I\approx  0.56$ mT/A.

This configuration with opposite electric fields on the two arms,
proposed by Wei et al. \cite{wei95}, differs from the idea of He,
McKellar and Wilkens \cite{he93,wilkens94} in which the same electric
dipole propagates in opposite magnetic fields but we think that the
phase shift predicted by equation \ref{e2} is truly a HMW phase shift
\cite{lepoutreSM}. We have replaced the charged wire of Wei et al.
\cite{wei95} by plane capacitors to improve the field homogeneity and
to minimize the dispersion of the polarizability phase (a large
dispersion reduces the fringe visibility). With this interaction
region and lithium atoms \cite{miffre06}, the HMW phase shift
predicted by equation (\ref{e2}) is
$\varphi_{HMW}(V,I)\mbox{(rad)}=-1.28\times10^{-6} VI$, corresponding
to a maximum value  $\left|\varphi_{HMW}\right|_{max}=27$ mrad for
$\left|V\right|=800$ V and $\left|I\right|=25$ A.

Let $\varphi_{E+B}(V,I)$ be the measured phase shift when the
electric and magnetic fields are both applied. $\varphi_{E+B}$
includes the HMW phase shift but also the dynamic phase shifts
\cite{cronin09} induced by each field acting separately. The electric
field induces a Stark (or polarizability) phase shift
\cite{ekstrom95} $\varphi_{S}=-2\pi\varepsilon_0 \alpha \oint
\mathbf{E}^2 dt/\hbar $. We tune the ratio of the voltages applied to
the capacitors to ensure $\varphi_{S} \lesssim 100$ mrad, a very
small value compared to the phase shifts induced on each arm which
can exceed $300$ rad.

The magnetic field induces a Zeeman energy shift $U_Z(F,m_F)$,
function of the $F,m_F$ hyperfine sub-level, and a phase shift
\cite{schmiedmayer94,jacquey07} given by  $\varphi_Z(F,m_F)=\oint
U_Z(F,m_F,B)dt/\hbar$. If a magnetic field $B=14$ mT was applied to
one arm only, the Zeeman phase shift would be extremely large,
$\varphi_Z \left(F = 2, m_F =\pm 2 \right) \approx  \pm 10^5$ rad.
This phase shift would be perfectly canceled if the magnetic field
had exactly the same value on the two arms but a weak field gradient
exists near the septum, sufficient to induce a Zeeman phase shift
$\varphi_Z \left(F = 2, m_F = \pm 2 \right) \approx \pm 11$ rad when
$B=14$ mT.  The detected signal is an average over the ground state
sub-levels with almost equal contributions
\cite{lepoutreSM,jacquey07}. These sub-levels form $4$ pairs with
opposite Zeeman phase shifts, so that the measured Zeeman phase shift
$\varphi_B$ is very weak but the large dispersion of $\varphi_Z$ with
$F,m_F$ reduces the fringe visibility. To reduce the Zeeman phase
shifts, another coil producing an opposite magnetic field gradient
was introduced outside the interaction region of fig. \ref{fig1}b,
and this compensation is excellent when hyperfine uncoupling is
negligible \cite{lepoutreSM}.

In order to extract the HMW phase, we perform measurements with
the voltage only to get the polarizability phase $\varphi_E(V)$
and with the current only to get the Zeeman phase  $\varphi_B(I)$, and
we then get $\varphi_{EB}(V,I)$:

\begin{equation} \label{e4}
 \varphi_{EB}(V,I)= \varphi_{E+B}(V,I)- \varphi_E(V)- \varphi_B(I)
\end{equation}

\noindent The diffraction phase $\varphi_d$ is very sensitive to the
$x$-positions of the standing wave mirrors M$_i$, with
$\delta\varphi_d/\delta x_i \approx 20$ rad/$\mu$m  for M$_1$, M$_3$
or $40$ rad/$\mu$m for M$_2$. This sensitivity induces phase drifts,
near $2$ rad/hour, due to small displacements $\delta x_i$ of thermal
origin. To minimize the effect of these drifts, we alternate
voltage-current configurations over a $20$ second-long fringe scan,
during which the phase drift is linear and it only modifies slightly
the scan slope. We have used either a 4-field configuration with the
following $(V,I)$ values $(0,0)$, $(V,0)$, $(V,I)$, $(0,I)$ or a
6-field configuration including $E$-reversal, by adding $(-V,0)$,
$(-V,I)$. Fits extract the characteristics of the individual fringe
patterns (see fig. \ref{fig2}). If  $\varphi(V,I)$ is the fringe
phase of the $(V,I)$ configuration, the phases needed to evaluate
equation \ref{e4} are given by the following differences,
$\varphi_E(V)= \varphi(V,0)- \varphi(0,0)$,  $\varphi_B(I)=
\varphi(0,I)- \varphi(0,0)$ and $\varphi_{E+B}(V,I)= \varphi(V,I)-
\varphi(0,0)$. The statistical uncertainty on $\varphi_{EB}(V,I)$,
near $30$ mrad for a $20$ second-long scan, is reduced near $3$ mrad
by averaging about $100$ scans.

\begin{figure}[t]
\begin{center}
\includegraphics[height= 6 cm]{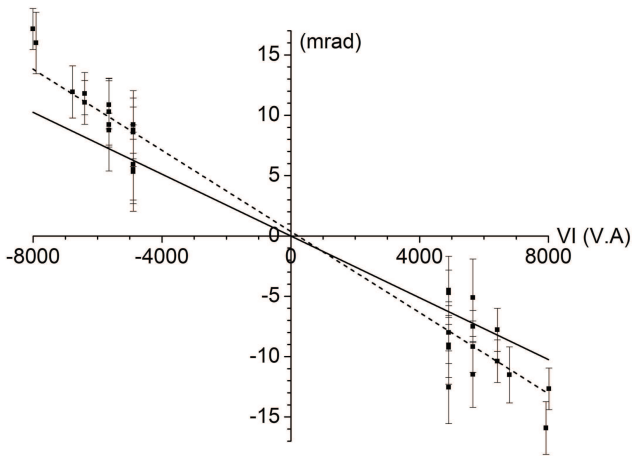}
\caption{He-McKellar-Wilkens phase shift as a function of the
voltage-current product: the phase $\varphi_{final}(V,I)$ is plotted
as a function of the $VI$ product. The data points (squares), represented with their
statistical error bars, are in good agreement with a linear behavior
$\varphi_{final}(V,I)\propto  VI$, as shown by the best fit to the
data (dotted line). The fit slope $(-1.68 \pm 0.07)×10^{-6}$  rad/V.A
is $31$\% larger than the expected slope of the HMW phase (full
line). \label{fig3}}
\end{center}
\end{figure}

The phase shift thus deduced is still influenced by various stray
phase shifts due to geometrical defects of the interaction region
\cite{lepoutreSM}. These systematic effects have been studied for a
large set of $(V,I)$ data points and an approximate analytical model was developed in order to evaluate their contributions to $\varphi_{EB}$. As the stray phase shifts increase rapidly with $I$, we discuss here only the data collected with
$\left|I\right| \leq 12$ A, but the experiments carried with larger
$\left|I\right|$-values have been useful to understand experimental
defects. The dominant part of the stray phase shifts is an even
function of $I$ and we separate it from $\varphi_{HMW}$ which
is odd with $V$ and $I$ by combining measurements with opposite
$I$-values in $\varphi_{final} =[\varphi_{EB}(V,I)-
\varphi_{EB}(V,-I)]/2$. Similarly as for the Zeeman phase shifts,
cancelations between hyperfine sublevels limit the contribution of
the Aharonov-Casher effect to very small values in $\varphi_{EB}$;
this contribution, always smaller than $3$ mrad, was evaluated thanks
to our model and subtracted from the data plotted in fig. \ref{fig3}.
Our results agree with the expected linear dependence $\varphi_{HMW}
\propto VI$, but the slope $\varphi_{final}(V,I)/(VI)=(-1.68 \pm
0.07)\times10^{-6}$ rad/(V.A) differs by $31$\% from the expected
value, presumably due to a lack of accuracy of our analytical model
\cite{lepoutreSM}.

The magnetic field direction changes over the interferometer, thus
inducing a Berry's phase \cite{shapere89}. The measured Berry's
phase is the difference between the two arms and it is expected to be
very small because of the magnetic field homogeneity. Moreover,
it is canceled by our procedure because it has the same
value in $\varphi_{E+B}(V,I)$ and in $\varphi_B(I)$. The more complex
Berry's phase involving the electric and magnetic fields discussed in
\cite{abdullah90} also appears to be negligible \cite{lepoutreSM}.

In conclusion, we have performed the first experimental test of the
He-McKellar-Wilkens topological phase by atom interferometry: this
experiment has taken advantage of the high phase sensitivity of our
interferometer and of its arm separation. Our measurement is not
limited by the interferometer sensitivity but by systematic effects
due to several small experimental defects.  The slope of the HMW
phase $\varphi_{final}(V,I)$ as a function of the $VI$ product
differs from the theoretical value and this difference is 
probably due to a lack of perfect understanding of the corrections
due to field gradients and other experimental defects. These defects
are small but they could be further reduced by a better design of the
interaction region. Moreover, the stray phase shifts due to these
defects are enhanced by the fact that the signal is an average over $8$
hyperfine-Zeeman sublevels and optical pumping in a single $F,m_F$
sub-level should greatly reduce these systematic effects and improve
the accuracy of the measurements. The HMW phase is expected to
independent of the atom velocity v and we can test this property by
tuning  lithium velocity by changing the supersonic beam carrier gas \cite{lepoutreSM}.
This test is feasible and very interesting but we must first reduce
the error bar in order to  distinguish this behavior from the
1/v-dependence of dynamic phases.

This experiment continues the development of new tools in quantum manipulation of atoms. For instance, the HMW effect could be used to build a coherent atom diode based on the non-reciprocal feature of topological phases. While existing atom diodes  \cite{ruschaupt06,thorn08} use optical pumping and do not conserve the coherence of the atom wave, the HMW effect is coherent. In an interferometer as shown in fig. \ref{fig1}, with a HMW phase  $\varphi_{HMW} =\pi/2$ for left to right propagation and a diffraction phase  $\varphi_d=-\pi/2$, a wave packet would be fully transmitted from entrance A to exit C but a wave packet entering by C would be transmitted to B and not to A. Although very intriguing, such a behavior does not violate any fundamental law and cannot be used to build a Maxwell demon \cite{maxwell75}.

We thank the laboratory technical staff for their help and A. Cronin for fruitful discussions. The development of our atom interferometer owes much to A. Miffre and to M. Jacquey who also made the first plans of the HMW experiment. We thank CNRS INP, ANR (grants ANR-05-BLAN-0094 and ANR-11-BS04-016-01 HIPATI) and R\'egion Midi-Pyr\'en\'ees for support.


\end{document}